\documentclass{article}

\usepackage{arxiv}

\usepackage[utf8]{inputenc} 
\usepackage[T1]{fontenc}    
\usepackage{hyperref}       
\usepackage{url}            
\usepackage{booktabs}       
\usepackage{amsfonts}       
\usepackage{nicefrac}       
\usepackage{microtype}      
\usepackage{lipsum}
\usepackage{graphicx}
\usepackage{dcolumn}
\usepackage{bm}
\usepackage{epsfig}
\usepackage{subfig}
\usepackage{amsmath}

\graphicspath{ {./images/} }

\DeclareMathOperator{\tr}{tr}

\def\Re {\mathbb{R}}

\title{Tight bounds of quantum speed limit for noisy dynamics via maximum rotation angles}

\author{
Zihao Hu \\
School of Intelligence Science and Engineering\\ 
Harbin Institute of Technology\\ 
Shenzhen 518055, China\\
Department of Mechanical and Automation Engineering\\
 The Chinese University of Hong Kong\\
 Shatin, Hong Kong SAR, China \\
  \texttt{zhhu2409@gmail.com} \\
   \And
Haidong Yuan \\
 Department of Mechanical and Automation Engineering\\
 The Chinese University of Hong Kong\\
 Shatin, Hong Kong SAR, China \\
  \texttt{hdyuan@mae.cuhk.edu.hk} \\
  \And
Zigui Zhang \\
School of Intelligence Science and Engineering\\ 
Harbin Institute of Technology\\ 
Shenzhen 518055, China\\
  \texttt{cnton10@foxmail.com} \\
  \And
Chi-Hang Fred Fung\\
Center for Computational Simulation\\
Universidad Politécnica de Madrid\\
Madrid 28040, Spain\\
  \texttt{chffung.app@gmail.com} \\
  \And
  Zibo Miao \\
School of Intelligence Science and Engineering\\ 
Harbin Institute of Technology\\ 
Shenzhen 518055, China\\
  \texttt{miaozibo@hit.edu.cn} \\
}

\begin{document}
\maketitle
\begin{abstract} {The laws of quantum physics place a limit on the speed of computation. In particular, the evolution time of a system from an initial state to a final state cannot be arbitrarily short.
 Bounds on the speed of evolution for unitary dynamics have long been studied. A few bounds on the speed of evolution for noisy dynamics have also been obtained recently, which are, however, not tight. In this paper, we present a new framework for quantum speed limit concerning noisy dynamics. Within this framework, we obtain the exact maximum rotation angle that noisy dynamics can achieve at any given time, which gives rise to a tight bound on the evolution time for noisy dynamics. The bound obtained through semi-definite programming highlights the fundamental differences between noisy dynamics and unitary dynamics. Furthermore, we show that the \textit{orthogonalization} time, defined as the minimum time required to evolve any initial state to a state with zero fidelity with respect to the initial state, is generally not applicable to noisy dynamics.}
\end{abstract}
\keywords{quantum speed limit, noisy dynamics}

\section{Introduction}

Quantum information processing can be regarded as the transformation of quantum states that encode the information to be processed or computed.
The time for which the states transform dictates the speed of the quantum computation. Quantum physics imposes a limit on the transformation time. This quantum speed limit (QSL)~\cite{Lloyd} arises because the energies of the system as well as the environment are finite and the state of the system may evolve according to slow dynamics. During a period of time $t$, a quantum process can rotate a quantum state by the angle $\theta$. In terms of QSL, the reverse question is asked. Namely, given a certain angle $\theta$, we ask what minimum time $t$ is required to rotate any state by angle $\theta$.

The first major result of QSL, which was based on the uncertainty relation, was made by Mandelstam and Tamm~\cite{Madelstam} in 1945. Since then, there has been an interest and development in the topic of QSL, including generalization to mixed states, Markovian and non-Markovian dynamics, closed and open quantum systems, different targets such as gauge invariant distances and Bloch angles, and many other applications including control strategies and shortcuts to adiabaticity associated with QSL~\cite{Pintos2022,ness2022,campo2021,ness2021,becker2021,hu2020,girolami2019,cai2017,camplbell2017,MB2017,deffner2017,Beau2019,chenu2017,pires2016,sun2021,sun2019,campaioli2019, campaioli2018,sxwu2018,marvian2015,sun2015,giova2004,SLZYL2023,SLZYL2021,SLZYL2020,YuanUniversal,Fung3,Fung2,Fung1,Chau2011,Xu20142,Xu2014,Sam2014,Zhang2014,Zhang2015,Liu2015,Deffner,Campo,Taddei,Judy2008,Borras2006,Batle2005,Giov2003,Vaidman,Fu2010,Luo2005,Luo2004,Caneva2009,Levitin2009,Brody2003,Giov03,Jones,Chau2010,Margolus,modal2016,LYAX1624,WA22,WA23}. Although various results on unitary dynamics have come out~(see e.g. \cite{Margolus,giova2004,Giov03,Chau2010,Jones, Levitin2009,Brody2003,Caneva2009, Luo2004, Luo2005, Fu2010, Vaidman, Giov2003, Batle2005,Borras2006,Judy2008}), studies on noisy dynamics and open quantum systems have only been carried out recently.
For example, QSL characterization schemes have been enriched in \cite{WA23} for open systems, particularly for addressing non-Markovian dynamics.

In this paper, we present a new framework for QSL concerning noisy dynamics. Although previous studies mostly focus on the rotating speed of a given state under certain dynamics, here we study the maximal speed of evolution that the dynamics can generate on all quantum states, which requires optimization over all states. The obtained speed of evolution represents the limit of quantum speed that the given dynamics can possibly induce on any quantum states, which is then a fundamental limit of the dynamics and can be used to provide bounds on the computation speed of a quantum device. While the QSL on a fixed state tells little about the ability of the dynamics with regard to rotation of the states in general, the maximal speed of evolution provides a way to gauge the dynamics.

Our framework is based on a method that gives the exact maximum rotation angle for certain given dynamics, which ensures that the bound is achievable. The bound is obtained directly from the Kraus operators of the dynamics, allowing for the ease of computation. The bound obtained reveals that noisy dynamics is essentially different from unitary dynamics. In particular, we show that the {\it orthogonalization} time, a concept commonly used in QSL, is in general not applicable to noisy dynamics.

Our framework is based on a distance measure in quantum channels, which will be briefly described in the following.
For an $m\times m$ unitary matrix $U$, we denote by $e^{-\mbox{i}\theta_j}$ the eigenvalues of $U$, where $\theta_j\in(-\pi,\pi]\ (1\leq j\leq m)$ is also referred to as the eigen-angles of $U$. We define (see, e.g. \cite{Chau2011,Fung1,Fung2})
$\parallel U\parallel_{\max}=\max_{1\leq j \leq m}\mid\theta_j\mid,$
and $\parallel U\parallel_g$ as the minimum of $\parallel e^{i\gamma}U\parallel_{\max}$ over equivalent unitary operators with different global phases, i.e. $\parallel U\parallel_g=\min_{\gamma\in \Re} \parallel e^{i\gamma} U\parallel_{\max}$. We then define
\begin{eqnarray}
    C_{\theta}(U)=\left\{\begin{array}{cc}
	\parallel U\parallel_g, & \mathrm{if} \parallel U\parallel_g\leq \frac{\pi}{2}, \\
	\frac{\pi}{2}, & \mathrm{if} \parallel U\parallel_g> \frac{\pi}{2}.\\
    \end{array}\right.
\end{eqnarray}
Essentially $C_{\theta}(U)$ represents the maximum angle at which $U$ can rotate a state away from itself \cite{Fung2}, that is,\\
\begin{eqnarray}
    C_{\theta}(U)=\arccos\min_{\rho}F_B(\rho, U\rho U^\dagger)
\end{eqnarray}
where the fidelity $F_B(\rho_1,\rho_2)$ between two states is defined as $F_B(\rho_1,\rho_2)= \tr \sqrt{\rho_1^{\frac{1}{2}}\rho_2\rho_1^{\frac{1}{2}}}$.  For an operator $X$, $X^\dagger$ denotes the adjoint of $X$. If the eigen-angles of $U$ are arranged in decreasing order, i.e. $\theta_{\max}=\theta_1\geq \theta_2\geq \cdots \geq \theta_m=\theta_{\min}$, then $C_{\theta}(U)={(\theta_{\max}-\theta_{\min})}/2$ when $\theta_{\max}-\theta_{\min}\leq \pi$ \cite{Fung2}.

Similarly, a distance metric $d(U_1,U_2)$ on unitary operators $U_1$ and $U_2$ can be induced by $C_{\theta}(\cdot)$ as
\begin{eqnarray}
    d(U_1,U_2)=C_{\theta}(U_1^\dagger U_2)=\arccos\min_{\rho}F_B(U_1\rho U_1^\dagger, U_2\rho U_2^\dagger).
\end{eqnarray}
The distance metric $d(U_1,U_2)$ represents the maximum angle that $U_1$ and $U_2$ can generate on the same input state $\rho$. This metric can be generalized to noisy dynamics as $d(K_1, K_2)=\min_{U_{ES_2}}d(U_{ES_1}, U_{ES_2}),$
where $U_{ES_1}$ and $U_{ES_2}$ are unitary extensions of Kraus opertors $K_1$ and $K_2$ respectively. 

To be concrete, for noisy dynamics, $d(K_1,K_2)$ represents the maximum angle that $K_1\otimes I_A$ and $K_2\otimes I_A$ can generate with respect to the same input state; The metric can be computed by 
\begin{eqnarray}
    d(K_1,K_2)=\arccos\min_{\rho_{SA}}F_B[K_1\otimes I_A (\rho_{SA}), K_2\otimes I_A(\rho_{SA})],
\end{eqnarray}
where $\rho_{SA}$ is a state of the composite systems consisting of the target and the ancilla, with $I_A$ denoting the identity operator defined in the ancillary system. 
Moreover, the metric $d(K_1,K_2)$ can also be obtained by
\begin{eqnarray}\label{eq:noisy_metric}
    d(K_1, K_2)=\arccos \max_{\|W\|\leq 1}\frac{1}{2}\lambda_{\min}(K_W+K^\dagger_W),
\end{eqnarray}
where $\lambda_{\min}(\cdot)$ denotes taking the minimum eigenvalue, and $K_W=\sum_{j=1}^D\sum_{i=1}^Dw_{ij}F_{1i}^\dagger F_{2j}$. Here $F_{1i}$ and $F_{2j}$, denote the Kraus operators of $K_1$ and $K_2$ respectively, $w_{ij}$ denotes the $ij$-th entry of a $D\times D$ matrix $W$ with $\|W\|\leq 1$
($\|\cdot\|$ is the operator norm which is equal to the maximum singular value).

Furthermore, denote $t=2\cos d(K_1,K_2)$, and such a distance can be efficiently calculated via semi-definite programming as
\begin{eqnarray}\label{eq:sdp}
    \aligned
    \max \qquad &\frac{1}{2}t \\
        \mathrm{s.t.}\qquad &\left(\begin{array}{cc}
	   I & W^\dagger  \\
	   W & I \\
        \end{array}\right)\succeq 0,\\
    & K_W+K^\dagger_W-tI \succeq 0.
    \endaligned
\end{eqnarray}
And the corresponding dual semi-definite programming provides a way to find the optimal state:
\begin{eqnarray}\label{eq:SDPrho}
    \aligned
    \min \qquad &\frac{1}{2}\tr(P)+\frac{1}{2}\tr(Q) \\
        \mathrm{s.t.}\qquad &\left(\begin{array}{cc}
	   P & M^\dagger(\rho_S)  \\
	   M(\rho_S) & Q \\
        \end{array}\right)\succeq 0,\\
    & \tr (\rho_S)=1,\rho_S\succeq 0,\\
    \endaligned
\end{eqnarray}
where $P, Q$ are Hermitian matrices and $M(\rho_S)$ is a $D\times D$ matrix with its $ij$-th entry equaling $\tr(\rho_S F_{1i}^\dagger F_{2j})$. The optimal state is any pure state $\rho_{SA}$ with $\tr_A({\rho_{SA}})=\rho_S$, where $\rho_S$ is obtained from the above semi-definite programming.

The metric can be used to obtain a saturable bound for QSL. More precisely, for the dynamics $K_t(\rho)=\sum_iF_i(t)\rho F_i^\dagger(t)$, suppose that it takes $t$ units of time for the dynamics to rotate a state, possibly entangled with an ancillary system, with an angle $\theta$. Then $\theta= \arccos F_B[\rho_{SA}, K_t\otimes I_A(\rho_{SA})]\leq d(I, K_t)$, and thus a lower bound on the minimum time can be obtained by this inequality where the equality can be saturated when $\rho_{SA}$ takes the optimal input state. When $\rho_{SA}$ is restricted to separable states, the maximal rotation speed is reduced to the case without an ancillary system, which is in general slower. $d(I, K_t)$ thus provides a limit on the maximum angle that the given dynamics can generate on any state at the time $t$.

First of all, for unitary dynamics $U_t=e^{-\mbox{i}Ht}$, suppose it takes $t$ units of time to rotate a state $\rho$ with the angle $\theta \in [0,\frac{\pi}{2}]$. Then
$\theta
\leq d(I,U_t)
=\frac12{(E_{\max}-E_{\min})t},$
where $E_{\max}$ ($E_{\min}$) denotes the maximum (minimum) eigenvalue of $H$. The minimum time needed to rotate a state away with the angle $\theta$ is then bounded by $t\geq {2\theta}/{(E_{\max}-E_{\min})}$. 
This recovers previous results on the quantum speed limit for unitary dynamics \cite{Brody2003}. This bound is also known to be saturable with the input state $|\varphi\rangle=(|E_{\max}\rangle+e^{\mbox{i}\phi}|E_{\min}\rangle)/\sqrt2$, which can always rotate to an orthogonal state at the time $t=\pi/{(E_{\max}-E_{\min})}$. QSL bounded via the Bloch angle has been discussed in \cite{SLZYL2020}, while our framework generalizes this to arbitrary noisy dynamics through the metric \(d(K_1, K_2)\).

Here, $E_{\max}-E_{\min}$ can be seen as the energy scale of the system, and thus $d(I, U_t)$ is proportional to the multiplication of the energy scale and time. The maximum angle that can be rotated is thus proportional to the time-energy cost of the dynamics \cite{Chau2011,Fung1,Fung2,Fung3}. For noisy dynamics, such as $d(I, K_t)=\min_{U_{ES_t}}(I_{ES}, U_{ES_t})$ where $U_{ES_t}$ is the unit extension of $K_t$, the maximum angle is proportional to the minimum energy cost on all unit extensions of noisy dynamics \cite{Chau2011,Fung1,Fung2,Fung3}. Unlike the quantum Fisher metric, which depends on the specific dynamical trajectory and may overestimate the evolution time, our metric \( d(K_1, K_2) \) directly quantifies the worst-case rotation angle over all possible input states. This ensures a tight bound that is saturable by an optimal state, even in the presence of decoherence.

In the following part of this paper, we will focus on the analysis of QSL concerning noisy dynamics.

\section{QSL for single systems}\label{sec:QSLss}

In this section, we are concerned with the analysis of QSL, characterized by the maximum rotation angle, under noisy dynamics for single systems.

\subsection{Dynamics with amplitude damping}\label{subsec:dynm_ampl_damp}

Consider the Markovian dynamics with amplitude damping $K_t(\rho)=F_{11}(t)\rho F_{11}^\dagger (t)+F_{12}(t)\rho F_{12}^\dagger (t)$, where the Kraus operators are
\begin{eqnarray}
    F_{11}(t)=\begin{bmatrix}
	1 & 0\\
	0 & \sqrt{P(t)}\end{bmatrix},\quad
    F_{12}(t)=\begin{bmatrix}
	0 & \sqrt{1-P(t)}\\
	0 & 0
    \end{bmatrix}.
\end{eqnarray}
Here, the time-varying element $P(t)=e^{-\gamma t}$ with $\gamma$ being the decay rate. Suppose that it takes $t$ units of time for the dynamics to rotate a state $\rho_{SA}$ with angle $\theta\in [0,\frac{\pi}{2}]$. The density operator $\rho_{SA}$ represents the quantum state of the target system and the ancilla, and then one can have $\theta= \arccos F_B[\rho_{SA}, K_t\otimes I_A(\rho_{SA})]\leq \arccos \min_{\rho_{SA}}F_B[\rho_{SA}, K_t\otimes I_A(\rho_{SA})]=d(I, K_t)$.

One can have $\cos d(I, K_t)=\max_{\|W\|\leq 1}\frac{1}{2}\lambda_{\min} (K_W+K_W^\dagger)$, where $K_W=\sum_{ij}w_{ij}F_{0i}^\dagger F_{1j}$. with $F_{01}=I$ and $F_{02}=\mathbf{0}$ being the Kraus operators for the identity operator (where a zero operator has been added). Here $w_{ij}$ is the $ij$-th entry of the $2\times 2$ matrix $W$ satisfying $\|W\|\leq 1$. Then 
\begin{eqnarray}
    K_W+K_W^\dagger=\begin{bmatrix}
	a & c\\
	c^* & b
    \end{bmatrix}
    =\begin{bmatrix}
        2 \Re(w_{11}) & w_{12}\sqrt{1-P(t)}\\
        - & 2 \Re(w_{11})\sqrt{P(t)}
    \end{bmatrix},
\end{eqnarray}
where $\Re(\cdot)$ denotes the real part of a number. The minimum eigenvalue of $K_W+K_W^\dagger$ can thus be given by $\lambda_{\min}(K_W+K_W^\dagger)=\frac12\left({a+b-\sqrt{(a-b)^2+4|c|^2}}\right)$. To maximize the minimum eigenvalue, $c$ should be set to $0$. More precisely, by choosing $w_{12}=0$, the expression for the minimum eigenvalue becomes $\lambda_{\min}(K_W+K_W^\dagger)=b=2 \Re(w_{11})\sqrt{P(t)}$ which reaches its maximum value when $w_{11}=1$. Therefore, $\cos d(I, K_t)=\max_{\|W\|\leq 1}\frac{1}{2}\lambda_{\min} (K_W+K_W^\dagger)=\sqrt{P(t)}$.
 As $\theta\leq d(I,K_t)$, we have $\cos \theta \geq \cos d(I,K_t)=\sqrt{P(t)}$, which gives $t\geq \frac{2}{\gamma}\ln \sec\theta$. This provides a lower bound for the minimum time needed to rotate any state with the angle $\theta$, and it is consistent with the previous results (see, e.g. \cite{Taddei}). Please note that in this scenario, to rotate a state to its orthogonal state, infinite time is needed as $\ln \sec\frac{\pi}{2}\rightarrow \infty$. In fact, this corresponds to the case where the initial state is the excited state $|1\rangle$ and only completely decays to the ground state $|0\rangle$ in an infinite amount of time.
	
For non-Markovian dynamics, due to strong couplings with the environment, the decay rate $\gamma_{nM}(t)$, which is usually time-dependent, can be greater than the decay rate in the Markovian regime \cite{Deffner}. Therefore, in such a case $P(t)=e^{-\int_0^t \gamma_{nM}(\tau)d\tau}$ where $\int_0^t \gamma_{nM}(\tau)d\tau$ is usually larger than $\gamma t$ in the Markovian regime, thus for the same time duration the maximum angle $d(I,K_t)=\arccos \sqrt{P(t)}$ can be bigger in the non-Markovian regime than in the Markovian regime. This was explored in previous studies showing that non-Markovian dynamics can contribute to quantum speed up \cite{Deffner,LYAX1624,WA22}. Please note that even in the non-Markovian regime, as long as $\gamma_{nM}(t)$ is finite, it always takes an infinite amount of time for $P(t)$ to reach $0$. Thus, an infinite amount of time is needed to achieve a $\pi/2$-rotation.

\subsection{Dynamics with dephasing noise}\label{subsec:dynm_dephase}

Let $K_t(\rho)=F_{11}(t)\rho F_{11}^\dagger (t)+F_{12}(t)\rho F_{12}^\dagger (t)$ describe the dynamics in the presence of dephasing noise, with the Kraus operators 
\begin{eqnarray}\label{eq:daphase_Kraus_ops}
    F_{11}(t)=\sqrt{\frac{1+P(t)}{2}}\begin{bmatrix}
		e^{-\mbox{i}\omega t/2} & 0\\
		0 & e^{\mbox{i}\omega t/2}
	\end{bmatrix},\quad
    F_{12}(t)=\sqrt{\frac{1-P(t)}{2}}\begin{bmatrix}
		e^{-\mbox{i}\omega t/2} & 0\\
		0 & -e^{\mbox{i}\omega t/2}
	\end{bmatrix}.
\end{eqnarray}
Here, $P(t)=e^{-\gamma t}$ and $\gamma$ denote the dephasing rate. We similarly suppose that it takes $t$ units of time for the dynamics to rotate the quantum state $\rho_{SA}$ with the angle $\theta\in [0,\frac{\pi}{2}]$, and thus $\theta\leq d(I, K_t)$.
In this scenario, we have that $K_W+K_W^\dagger={\rm diag}[a,\ b]$ with
\begin{eqnarray}
    \left\{\begin{array}{l}
        a=2\Re\left(\sqrt{\frac{1+P(t)}{2}}w_{11}e^{-\mbox{i}\omega t/2}+\sqrt{\frac{1-P(t)}{2}}w_{12}e^{-\mbox{i}\omega t/2}\right)\\
        b=2\Re\left(\sqrt{\frac{1+P(t)}{2}}w_{11}e^{\mbox{i}\omega t/2}-\sqrt{\frac{1-P(t)}{2}}w_{12}e^{\mbox{i}\omega t/2}\right)
    \end{array}\right.
\end{eqnarray}
By using $|w_{11}|^2+|w_{12}|^2\leq 1$ for any $\|W\|\leq 1$ together with the Cauchy-Schwarz inequality, one can obtain that
\begin{equation}
    \begin{aligned}
        \lambda_{\min}(K_W+K_W^\dagger)\leq &\frac{1}{2}\tr(K_W+K_W^\dagger)\\
	   =&2\Re\left(\sqrt{\frac{1+P(t)}{2}}w_{11}\cos(\omega t/2)-\mbox{i}\sqrt{\frac{1-P(t)}{2}}w_{12}\sin(\omega t/2)\right)\\
        \leq &2\left(\left|\sqrt{\frac{1+P(t)}{2}}w_{11}\cos(\omega t/2)\right|+\left|\sqrt{\frac{1-P(t)}{2}}w_{12}\sin(\omega t/2)\right|\right)\\
        \leq &2\sqrt{\frac{1+P(t)}{2}\cos^2(\omega t/2)+\frac{1-P(t)}{2}\sin^2(\omega t/2)}
        \sqrt{|w_{11}|^2+|w_{12}|^2}\\
        \leq &2\sqrt{\frac{1+P(t)\cos(\omega t)}{2}}.
    \end{aligned}
\end{equation}
It is not difficult to verify that the equality is saturated when
\begin{eqnarray}\label{eq:dephase_W}
    W=\begin{bmatrix}
        \frac{\sqrt{1+P(t)}\cos(\omega t/2)}{\sqrt{1+P(t)\cos(\omega t)}}&\frac{\mbox{i}\sqrt{1-P(t)}\sin(\omega t/2)}{\sqrt{1+P(t)\cos(\omega t)}}\\
        0&0
    \end{bmatrix}
\end{eqnarray}

\begin{figure} 
    \centering
    \includegraphics[width=0.5\textwidth]{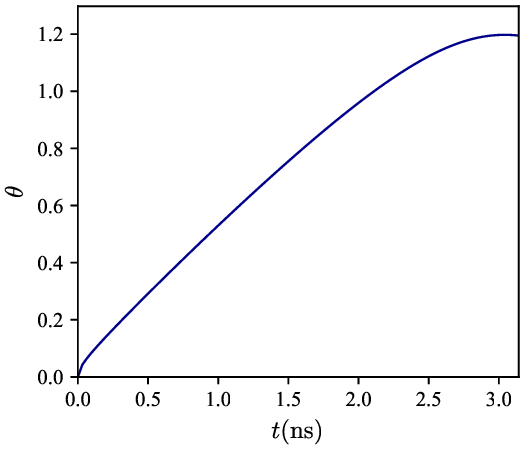}
    \caption{(Color online) The maximum angles that can be rotated at different values of the time $t$ in the presence of dephasing noise, with $\gamma=0.1~\text{GHz}$ and $\omega=1~\text{GHz}$.}
    \label{fig:bound}
\end{figure}

Then it can be concluded that $\cos d(I, K_t)=\sqrt{\frac{1+P(t)\cos(\omega t)}{2}}.$ And since $\cos\theta\geq \cos d(I,K_t)$, the minimum time needed to rotate a state with the angle $\theta$ can be obtained, as illustrated in Figure \ref{fig:bound}.

It is worth noting that $\cos d(I,K_t)=\sqrt{\frac{1+e^{-\gamma t}\cos(\omega t)}{2}}>0$ for $P(t)=e^{-\gamma t}$ as long as $\gamma> 0$. Hence $d(I,K_t)<\pi/2$; that is, the dynamics cannot rotate any state to its orthogonal state. 
This is a much stronger statement than the previous result in \cite{Taddei}, where it was stated that only when $\frac{\omega}{\gamma}> r_{crit}\approx 2.6$ the dynamics could not rotate any state to its orthogonal state.  This difference arises because the previous bound is obtained by integrating the quantum Fisher metric along the path $\rho_t=K_t\otimes I_A (\rho_{SA})$. This path is fixed by the dynamics, which is usually not the geodesic between the initial state and the final state. Consequently, the integration of the quantum Fisher metric along the path is in general larger than the actual distance between the initial state and the final state. This in turn leads to a looser bound and inaccurate classification for noisy dynamics. The bound obtained in \cite{Campo} for dynamics with dephasing noise is also not tight, which resulted in a finite {\it orthogonalization} time. In contrast, the bound obtained here is tight and can be saturated with the input state $|+\rangle=(|0\rangle+|1\rangle)/\sqrt2$. In addition, an ancillary system is not needed to saturate the bound we have obtained in the presence of dephasing noise.

\subsection{Generic noisy dynamics}\label{subsec:dynm_generic}

We will show that for generic noisy dynamics $K_t(\rho)=\sum_{i=1}^DF_i(t)\rho F_i^\dagger(t)$, if the identity operator $I$ belongs to the space spanned by the Kraus operators, then $K_t$ cannot rotate any state to its orthogonal state, or equivalently, $d(I, K_t)$ is always smaller than $\pi/2$. 

The reason lies in the fact that if $I\in {\rm span}\{F_1(t), F_2(t),\cdots, F_D(t)\}$, then there exists $w_{1i}$ such that $I=\sum_{i=1}^Dw_{1i}F_i(t)$. Now, let $\alpha=1/\sqrt{\sum_{i=1}^D |w_{1i}|^2}>0$, then $\alpha I= \sum_{i=1}^D w'_{1i}F_i(t)$ with $w'_{1i}=\alpha w_{1i}$. Define $W'$ as a matrix $D\times D$ with the entries of the first row equal to $w'_{1i}$ and the other entries equal to $0$. It is then obvious that $\|W'\|=1$, and thus
\begin{eqnarray}
    \label{eq:W'}
    \aligned
    \cos d(I, K_t)&= \max_{\|W\|\leq 1}\frac{1}{2}\lambda_{\min}(K_W+K_W^\dagger)\\
    &\geq \frac{1}{2}\lambda_{\min}(K_{W'}+K_{W'}^\dagger)\\
    &=\frac{1}{2}\lambda_{\min}\left[\sum_{i=1}^D w'_{1i}F_i(t)+\left(\sum_{i=1}^D w'_{1i}F_i(t)\right)^\dagger\right]\\
    &=\alpha>0.
    \endaligned
\end{eqnarray}
Hence $d(I, K_t)\leq \arccos \alpha <\pi/2$. That is, the dynamics cannot rotate any state to its orthogonal state.

For example, in the case of dephasing noise as indicated in Eq~(\ref{eq:daphase_Kraus_ops}),
$I=\sqrt{\frac{2}{1+P(t)}}\cos(\omega t/2)F_{11}(t)+\mbox{i}\sqrt{\frac{2}{1-P(t)}}\sin(\omega t/2)F_{12}(t)$, and then
\begin{eqnarray}\label{eq:alpha}
    \alpha=\frac{1}{\sqrt{\frac{2}{1+P(t)}\cos^2(\omega t/2)+\frac{2}{1-P(t)}\sin^2(\omega t/2)}}=\frac{\sqrt{1-P^2(t)}}{\sqrt{2-2P(t)\cos(\omega t)}},
\end{eqnarray}
which is positive for any $P(t)< 1$. Hence, in the presence of dephasing noise, $d(I,K_t)\leq\arccos\alpha<\pi/2$.

This fact can also be easily seen from the equivalent representations of the Kraus operators. More precisely, when $I\in {\rm span}\{F_1(t), F_2(t),\cdots, F_D(t)\}$, there exists an equivalent representation of Kraus operators such that $\alpha I$ is one of them. Then the fidelity between the initial and final states will be at least $\alpha$, and thus this dynamics cannot rotate any state to its orthogonal state. The bound proposed by us can not only reflect this fact, but can also provide a tighter bound by exploring different choices of $W$. Taking dynamics with dephasing noise, for example, the choice of $W$ in Eq~(\ref{eq:dephase_W}) can lead to a tight bound. In addition, it is not difficult to observe that if the span of Kraus operators contains any matrix $M$ such that $\lambda_{\min}(M+M^\dagger)>0$, the above argument holds. Thus, the dynamics cannot rotate any state to its orthogonal state. Taking dynamics with amplitude damping for example, the span of the associated Kraus operators contains $M={\rm diag}[1,\ \sqrt{P(t)}]$ which satisfies the condition $\lambda_{\min}(M+M^\dagger)=2\sqrt{P(t)}>0$ except for $P(\infty)=0$. 

An immediate implication is that all dynamics with the associated Kraus operators that span the entire space (or equivalently, the number of linearly independent Kraus operators is $d=n^2$, where $n$ denotes the dimension of the quantum system) cannot rotate any state to its orthogonal state. Such dynamics are indeed generic among all completely positive trace-preserving maps, therefore generic noisy dynamics cannot rotate any state to its orthogonal state.

\section{QSL for composite systems}\label{sec:QSLCS}

As discussed in Section~\ref{subsec:dynm_generic}, we now assume that there are $N$ numbers of such dynamics, denoted by $K_t^{\otimes N}$, acting independently in a composite system. The representation of the Kraus operators for $K_t^{\otimes N}$ can be written as $\tilde{F}_{i_1,i_2,\cdots, i_N}(t)=F_{i_1}(t)\otimes F_{i_2}(t)\otimes\cdots \otimes F_{i_N}(t)$. For the matrix $W'$ already discussed in Section~\ref{subsec:dynm_generic}, let $\tilde{W}=W'^{\otimes N}$, then $K^{\otimes N}_{\tilde{W}}=(K_{W'})^{\otimes N}=\alpha^N I^{\otimes N}$. One can thus have that
\begin{eqnarray}
    \aligned
    \cos d(I^{\otimes N}, K_t^{\otimes N})&=\max_{\|W\|\leq 1}\frac{1}{2}\lambda_{\min}(K^{\otimes N}_W+(K^{\otimes N}_W)^\dagger)\\
    &\geq \frac{1}{2}\lambda_{\min}\left(K^{\otimes N}_{\tilde{W}}+(K^{\otimes N}_{\tilde{W}})^\dagger\right)\\
    &=\lambda_{\min}(\alpha^N I^{\otimes N})\\
    &=\alpha^N>0,
    \endaligned
\end{eqnarray}
which implies $d(I^{\otimes N}, K_t^{\otimes N})\leq \arccos(\alpha^N)<\pi/2$.
It can then be concluded that in this case any state of the composite system cannot be rotated to its orthogonal state.

\begin{figure}
    \centering
    \includegraphics[width=0.9\textwidth]{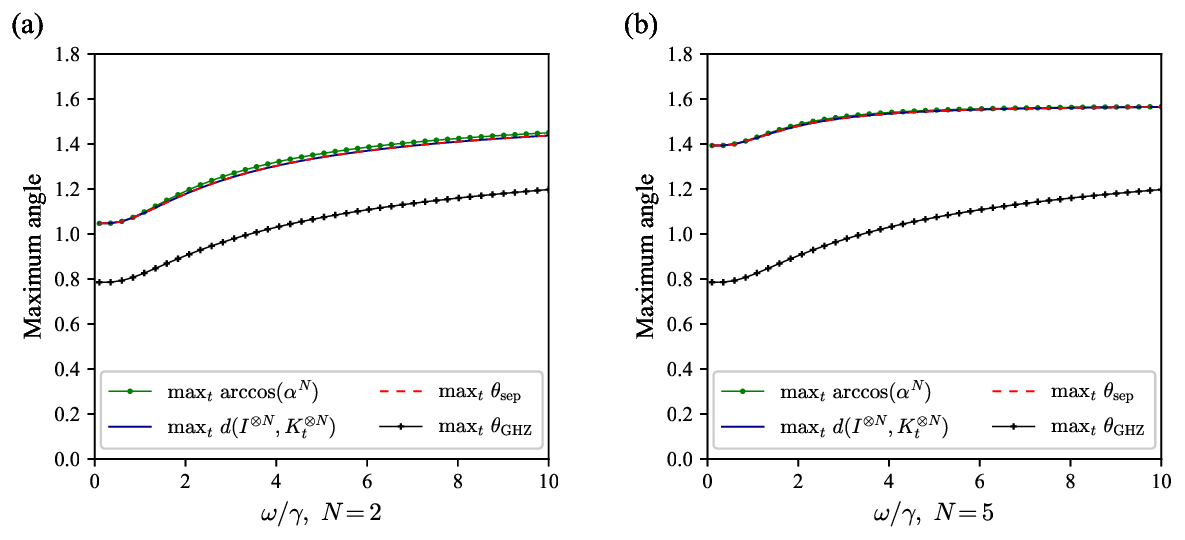}
    \caption{(Color online) The maximum angles that can be rotated  in composite systems in the  presence of dephasing noise, with the bounds at different values of $\frac{\omega}{\gamma}$ plotted. These curves are obtained by figuring out the optimal time spot $t$ that gives the maximum angle for the separable state, $d(I^{\otimes N}, K_t^{\otimes N})$ and $\arccos(\alpha^N)$, with (a) $N=2$ and (b) $N=5$ respectively. It can be seen that $\max_t\theta_{\text{sep}}$ and $\max_t\arccos(\alpha^N)$ provides tight bounds for $\max_t d(I^{\otimes N}, K_t^{\otimes N})$. The maximum angle that can be achieved with the GHZ state is also plotted for comparison.}
    \label{fig:maximalangle5}
\end{figure}

In fact, in the presence of dephasing noise, substituting the value of $\alpha$ into Eq.(\ref{eq:alpha}), one can obtain an upper bound for $d(I^{\otimes N}, K_t^{\otimes N})$ straightforwardly. A lower bound for $d(I^{\otimes N}, K_t^{\otimes N})$ can also be obtained by taking the input state as the separable state $|+\cdots+\rangle$, where $|+\rangle=(|0\rangle+|1\rangle)/\sqrt{2}$. It is then not difficult to calculate the rotated angle with respect to this separable state, which is $\theta_{\text{sep}}=\arccos (\beta^N)$ with $\beta=\sqrt{\frac{1+e^{-\gamma t}\cos(\omega t)}{2}}$, and thus
$\arccos (\beta^N)\leq d(I^{\otimes N}, K_t^{\otimes N})\leq \arccos(\alpha^N).$ Then the inequality $\max_t\arccos (\beta^N)\leq \max_t d(I^{\otimes N}, K_t^{\otimes N})\leq \max_t \arccos(\alpha^N)$ limits the maximum angle that can be rotated for composite systems. In Figure \ref{fig:maximalangle5}, we plot these bounds and the exact maximum angle for composite systems in the presence of dephasing noise for $N=2$ and $N=5$. It can be seen that these bounds are quite tight.

\begin{figure}[htbp]
    \centering
    \includegraphics[width=0.85\textwidth]{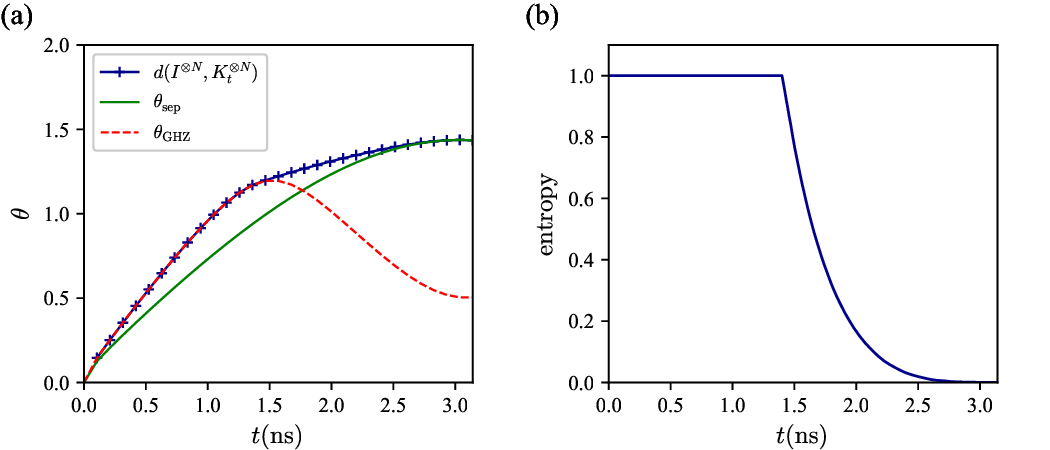}
    \caption{(Color online) Understanding QSL under the noisy dynamics for composite systems in the presence dephasing noise with the parameters chosen as $\gamma=0.1~\text{GHz}, \omega=1~\text{GHz}, N=2$. (a) Rotation angles on the GHZ and separable states respectively, compared with the maximum angles. (b) Quantified entanglement of the optimal input state which achieves the maximum rotation angle.}
    \label{fig:maximalangle2}
\end{figure}

On the other hand, for composite systems, the GHZ state (that is,$(|0\cdots0 \rangle+|1\cdots1\rangle)/\sqrt{2}$) is usually used as a benchmark for the QSL \cite{Taddei,Campo}. 
The rotation angle on the GHZ state can be explicitly computed as
$\cos\theta_{\text{GHZ}}=\sqrt{\frac{1+e^{-N\gamma t}\cos(N\omega t)}{2}}.$
It can be seen from Figure~\ref{fig:maximalangle2}(a) that for small values of $t$ (i.e., when the noise influence is still not strong), the GHZ state can help achieve the maximal speed of evolution. However, for high values of $t$, the GHZ state is no longer the optimal state that achieves the maximum angle $d(I^{\otimes N}, K_t^{\otimes N})$. More precisely, the GHZ state can be even worse than the separable state. This can be clearly observed in Figure~\ref{fig:maximalangle2}(b), where we quantify the entanglement for the optimal state that saturates $d(I^{\otimes^2}, K_t^{\otimes 2})$.

The maximally entangled state is optimal only when $t$ is below the threshold (e.g. $t < 1.5$). When $t$ is above the threshold, the optimal state that achieves the maximum rotation angle gradually changes from the maximum entangled state to the separable state. Moreover, it can be seen that the maximum angle on the GHZ state is much smaller than the maximum angle on the separable state. This is because the maximum angle on the GHZ state does not change with $N$, which can be observed from Figure \ref{fig:maximalangle5} (it only shortens the optimal time consumed to obtain the maximum angle by $N$ times). That is, $\max_t \theta_{\text{GHZ}}=\max_t\arccos \beta$ with $\beta=\sqrt{\frac{1+e^{-\gamma t}\cos(\omega t)}{2}}$, while $\max_t \theta_{\text{sep}}=\max_t\arccos (\beta^N)$ increases with $N$. From another perspective, if we take the rotated angle as the degenerate effect under noisy dynamics, it indicates that although the GHZ state deteriorates fast in the presence of dephasing noise in a short period of time, in the long run, the entanglement in the GHZ state mitigates the maximal degeneration.

\section{Conclusions and future work}
\label{sec:Con}
We provide a new framework to calculate tight bounds for QSL quantified by the exact maximum rotation angles under generic noisy dynamics (including non-Markovian dynamics). In particular, for arbitrary finite-dimensional quantum systems, the Kraus operators can be substituted in Eq.~\eqref{eq:SDPrho} to compute the maximum rotation angle. Similarly, composite systems of \( N \)-qudits or hybrid systems can be accommodated by the tensor products of Kraus operators. This generality ensures that our results are not restricted to specific dimensions or noise types. The maximum rotation angles and the corresponding bounds given in this paper clearly show that the commonly used concept for QSL, i.e. the {\it orthogonalization} time, is in general not applicable to noisy dynamics. The derived bounds obtained through semi-definite programming are achievable by the optimal input state, quantifying the fundamental limit imposed by the dynamics itself, which is critical for assessing the intrinsic capabilities of quantum processes. It is also shown that although maximally entangled states, such as the GHZ state, evolve faster in a short period of time, they are not optimal states, giving rise to maximum rotation angles under noisy dynamics in the long run. 

Furthermore, our work has significant implications for quantum computing, since the state transformation time bounds the speed of computation. Additionally, the amount of state degradation is bounded by the storage time, which in turn enhances our understanding of quantum memory.

\section*{Acknowledgments}
 The authors thank Jing Liu, Lingna Wang and Hongzhen Chen for helpful discussions. Zihao Hu and Haidong Yuan contributed equally to this work, and part of this work was done while Zihao Hu was visiting Harbin Institute of Technology (Shenzhen Campus).

\end{document}